\documentclass[prd,superscriptaddress,twocolumn,showpacs,amsmath,%
preprintnumbers,showkeys]{revtex4}
\usepackage{bm}
\usepackage{graphicx}
\usepackage{url}
\newcommand{\beq}{\begin{equation}}
\newcommand{\eeq}{\end{equation}}
\newcommand{\be}{\begin{eqnarray}}
\newcommand{\ee}{\end{eqnarray}}
\begin{document}                                           
\title{Realizing vector meson dominance with transverse charge densities}
\author{G.~A.~Miller}
\affiliation{Department of Physics, University of Washington, 
Seattle, WA 98195--1560, USA}
\author{M.~Strikman}
\affiliation{Department of Physics, Pennsylvania State University,
University Park, PA 16802, USA}
\author{C.~Weiss}
\affiliation{Theory Center, Jefferson Lab, Newport News, VA 23606, USA}
\date{May 25, 2010}
\begin{abstract}
The transverse charge density in a fast--moving nucleon 
is represented as a dispersion integral of the imaginary part of the 
Dirac form factor in the timelike region (spectral function). 
At a given transverse distance $b$ the integration effectively extends 
over energies in a range $\sqrt{t} \lesssim 1/b$, with exponential 
suppression of larger values. The transverse charge density at
peripheral distances thus acts as a low--pass filter for the spectral 
function and allows one to select energy regions dominated by specific
$t$--channel states, corresponding to definite exchange mechanisms 
in the spacelike form factor. We show that distances 
$b \sim 0.5 - 1.5 \, \textrm{fm}$ in the isovector density are maximally 
sensitive to the $\rho$ meson region, with only a $\sim 10\%$ contribution 
from higher--mass states. Soft--pion exchange governed by chiral dynamics 
becomes relevant only at larger distances. In the isoscalar density 
higher--mass states beyond the $\omega$ are comparatively more
important. The dispersion approach suggests that the positive transverse 
charge density in the neutron at $b \sim 1\, \textrm{fm}$, found previously 
in a Fourier analysis of spacelike form factor data, could serve as a 
sensitive test of the the isoscalar strength in the 
$\sim 1\, \textrm{GeV}$ mass region. In terms of partonic structure, 
the transverse densities in the vector meson region 
$b \sim 1 \, \textrm{fm}$ support an approximate mean--field picture 
of the motion of valence quarks in the nucleon.
\end{abstract}
\keywords{Vector meson dominance, electromagnetic form factors, 
dispersion relations, generalized parton distributions}
\pacs{12.40.Vv, 13.40.Gp, 11.55.Fv, 13.60.Hb}
\preprint{NT@UW-11-08, JLAB-THY-11-1377}
\maketitle
\section{Introduction}
Elastic electron scattering is one of the principal sources of information 
on the nucleon's spatial size and its internal structure. Two different 
physical pictures have traditionally been invoked to interpret the nucleon
form factors measured in such experiments. The first imagines the nucleon 
as an extended object in space, characterized by a distribution of charge 
and current, and aims to explain the form factors 
as the Fourier image of these spatial distributions. 
This approach has been used extensively in non-relativistic nuclear 
physics, where electron scattering has provided detailed spatial images 
of the charge and current distribution in nuclei. The other picture 
views elastic scattering as the exchange of a meson--like system between 
the current and the nucleon and attempts to describe the form factors 
in terms of the masses and couplings of these hadronic states. 
Historically, the existence of vector mesons was first postulated in 
order to explain the observed behavior of the nucleon form factors in the
region of spacelike momentum transfers $|t| \lesssim 1\, \textrm{GeV}^2$
\cite{Bernstein:1968}.
The equivalence of the ``extended object'' and ``exchange mechanism'' 
viewpoints is rooted in fundamental properties of strong interactions, 
namely their relativistic invariance and causality. 
They guarantee the existence of dispersion relations that express 
the form factors at spacelike momentum transfers in terms of their 
imaginary parts in the timelike domain (or spectral functions), 
where the exchange mechanisms correspond to intermediate hadronic 
states in the hypothetical process of nucleon--antinucleon creation
by the electromagnetic current.

It is generally expected that a more quantitative comparison between 
the two pictures might provide useful insights into nucleon structure. 
Generally, one hopes that in this way one may relate the physical
density of charge and current at a given distance to exchange 
mechanisms of a certain mass. However, such studies were long rendered 
unattractive by the fact that the conventional spatial representation 
of form factors, in terms of three--dimensional spatial distributions 
in the Breit frame (zero energy transfer), 
is meaningful only for non-relativistic systems. These distributions have 
no proper density interpretation in the relativistic case 
\cite{Miller:2007uy,Miller:2010nz} and cannot 
be related to observables in processes other than elastic $eN$ scattering.
The Breit frame distributions produced by the well-known exchange 
mechanisms were studied in several works, but it has proved difficult
to interpret the results outside of this particular 
context \cite{Friedrich:2003iz,Hammer:2003qv,Crawford:2010gv}.

A new approach to this problem is possible with the concept of transverse 
densities \cite{Soper:1976jc}, whose properties were explored in a series 
of recent 
articles \cite{Miller:2007uy,Miller:2010nz,Strikman:2010pu,Venkat:2010by}. 
They are defined as 2--dimensional Fourier transforms of the elastic form 
factors and describe the distribution of charge and magnetization in the 
plane transverse to the direction of motion of a fast--moving system. 
In contrast to the Breit frame distributions, they
are proper densities and permit a spatial interpretation also for 
systems in which the motion of the constituents is essentially relativistic,
such as hadrons in QCD. In fact, the transverse densities are closely 
related to the parton picture of 
hadron structure in high--energy processes and correspond to 
a reduction of the generalized parton distributions (or GPDs)
describing the distribution of quarks/antiquarks with respect 
to longitudinal momentum and transverse 
position \cite{Burkardt:2000za,Diehl:2002he}.
As such, they have an objective meaning beyond low--energy elastic 
$eN$ scattering and can be related to observables in certain
high--energy deep--inelastic processes sensitive to the transverse 
sizes of the nucleon, such as exclusive and diffractive $eN$ 
and $NN$ scattering \cite{Frankfurt:2005mc}. This places
the study of transverse densities in the wider context of exploring
the nucleon's partonic structure and allows one to employ concepts of
partonic dynamics to interpret the resulting spatial distributions.

In this article we study the transverse charge densities in the
nucleon's periphery in a dispersion representation which reveals the 
connection between partonic structure and the exchange mechanisms 
acting in the nucleon form factors. This approach was used previously 
to obtain the chiral large--distance component of  the charge density 
from a theoretical calculation of the isovector spectral function near 
threshold \cite{Strikman:2010pu}. Here we perform a more extensive
analysis using empirical spectral functions determined in an dispersion 
fit to nucleon form factor data \cite{Belushkin:2006qa}, which include
the vector meson region and the high--mass continuum and cover both
the isovector and isoscalar channels. Our study reveals several 
interesting aspects of the transverse charge densities. 

First, the transverse distance $b$ provides an external 
parameter which allows one to effectively select different energy 
(or mass) regions in the spectral function.
This happens thanks to the exponential convergence of the
dispersion integral for the transverse density, which strongly 
suppresses the contribution of energies $\sqrt{t} > 1/b$.
In particular, we show that distances $b \sim 0.5 - 1.5 \, \textrm{fm}$ 
maximally emphasize the $\rho$ meson mass region in the isovector 
spectral function, with only a $\sim 10\%$ contribution from 
higher--mass states. In the isoscalar channel the contribution
from higher--mass states above the $\omega$ are comparatively larger, 
but the $\omega$ can be isolated by going to larger distances 
of $\sim 2 \, \textrm{fm}$. The transverse densities at these distances 
represent, to our knowledge, the cleanest ``vector dominance'' 
observables, permitting detailed study of the vector meson couplings 
to the nucleon in spacelike (exchange) kinematics.
 
Second, the dispersion result for the transverse charge 
densities confirms a slightly positive density in the neutron at intermediate
distances $b \sim 0.5 - 1.5 \, \textrm{fm}$, found previously in a
Fourier analysis of the spacelike nucleon form 
factors \cite{Miller:2007uy}. While not unexpected --- the spectral 
functions were constructed to fit the spacelike form factor data --- 
this allows us to discuss this result from a $t$--channel perspective. 
The dispersion approach clearly shows 
that the ``pion cloud'' becomes relevant only at
distances $b \gtrsim 2 \, \textrm{fm}$, and that the positive
density at intermediate distances is dual to vector meson exchange,
with important contributions from higher--mass isoscalar states.
Their dynamical interpretation remains a challenging problem and
is related to the question of the strangeness content of the nucleon.
Measurements of the neutron charge density thus may be able to constrain
the couplings of these states to the nucleon.

Third, the dispersion results provide new
insight into the nucleon's partonic structure. By constructing the ratio
of $u$-- and $d$--quark transverse densities in the nucleon we show
that the ``vector dominance'' region $b \sim 1\, \textrm{fm}$
is consistent with an approximate mean--field picture of the motion of
valence quarks in the nucleon, as suggested by quark models.
Our approach allows us to formulate this duality in a 
model--independent manner, preparing the ground for dynamical
model studies.

The plan of this paper is as follows. In Sec.~\ref{sec:spectral} we 
discuss the basic properties of the dispersion representation
of transverse densities, focusing on the role of the 
distance $b$ as a filter for energies $\sqrt{t} \sim 1/b$ in the
spectral function. In Sec.~\ref{sec:isovector} we summarize present
knowledge of the isovector spectral function and study the contributions 
of the different energy regions to the transverse density, using the
empirical parametrization of Ref.~\cite{Belushkin:2006qa}. 
We identify the region of $\rho$ meson dominance and
quantify the corrections resulting from higher--mass states.
Appendix~\ref{app:theoretical} explains in detail how this analysis
relates to our earlier study of the chiral component of the isovector 
transverse density using chiral perturbation theory \cite{Strikman:2010pu}. 
In Sec.~\ref{sec:isoscalar} we consider the isoscalar charge density
and study its sensitivity to the $\omega$ meson pole in the 
spectral function. We also estimate its uncertainty at large $b$ 
and discuss at what momentum transfers
future measurements of the (spacelike) isoscalar form factor would 
have the strongest impact on the determination the $\omega NN$ coupling. 
In Sec.~\ref{sec:pn} we use our results to study the proton and 
neutron transverse charge density in the spectral representation.
In Sec.~\ref{sec:partonic} we extract the transverse densities
of $u$ and $d$ quarks in the dispersion approach and discuss the 
implications for the nucleon's partonic structure.
A summary and outlook are presented in Sec.~\ref{sec:summary}.

The dispersion representation can in principle be applied to study 
transverse densities at any distance, provided one has sufficient 
information on the relevant spectral functions. In this work we focus 
on the peripheral region of $b \gtrsim 0.5 \, \textrm{fm}$, where the 
densities are dominated by the low--mass singularities that are
well constrained by theoretical arguments and fits to present form 
factor data. While we use the parametrization of 
Ref.~\cite{Belushkin:2006qa} for our numerical studies, our conclusions 
are generic and rely on features that are common to all such approaches. 
Some recent form factor data that appeared after the fit of 
Ref.~\cite{Belushkin:2006qa} are incorporated in the discussions 
of Secs.~\ref{sec:pn} and \ref{sec:partonic} and support our 
numerical results. 
\section{Spectral representation of transverse densities}
\label{sec:spectral}
The matrix element of the vector current operator between nucleon
states with four--momenta $p_1$ and $p_2$ is parametrized by two 
functions of the invariant momentum transfer $t \equiv (p_2 - p_1)^2 < 0$,
the Dirac and Pauli form factors, $F_1(t)$ and $F_2(t)$; see 
Ref.~\cite{Perdrisat:2006hj} for conventions and basic properties. 
The Dirac form factor at zero momentum transfer 
is normalized to the total charge of the nucleon,
\beq
F_1^p(0) \;\; = \;\; 1, \hspace{2em} F_1^n(0) \;\; = \;\; 0. 
\eeq
Experimental knowledge of the nucleon form factors at spacelike 
momentum transfer is reviewed in Ref.~\cite{Perdrisat:2006hj};
for a discussion of the most recent data see e.g.\ Ref.~\cite{Cates:2011pz}.

%
%
\begin{figure}
\includegraphics[width=0.33\textwidth]{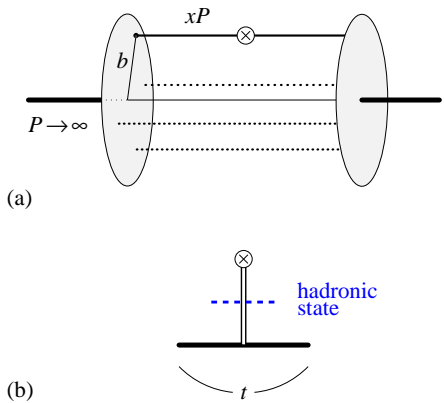}
\caption{(a) Partonic interpretation of the transverse charge density.
(b) Singularities of the timelike form factor resulting from transitions
to hadronic intermediate states.}
\label{fig:density}
\end{figure}
The transverse charge densities of the nucleon are defined as the 
two--dimensional Fourier transform of the Dirac form factors 
\be
\rho^{p, n} (b) &\equiv& \int \frac{d^2\Delta}{(2\pi)^2}
\; e^{-i (\bm{\Delta}\bm{b})}  \; F_1^{p, n}(t = -\Delta^2) 
\label{rho_fourier}
\\
&=& \int\limits_0^\infty \frac{d\Delta}{2\pi} 
\; \Delta \; J_0 (\Delta b) \; 
F_1^{p, n} (t = -\Delta^2 ) ,
\label{rho_fourier_radial}
\ee
where $\Delta \equiv |\bm{\Delta}|$ and $b \equiv |\bm{b}|$.
They have a simple interpretation in the infinite--momentum frame,
where the nucleon is moving fast in the ``longitudinal'' direction
and receives a momentum transfer $\bm{\Delta}$ in the ``transverse''
direction. In this frame the coordinate $\bm{b}$ measures the distance 
from the transverse center of momentum of the nucleon, and the
functions $\rho^{p, n}(b)$ describe the transverse spatial distribution 
of electric charge with normalization $\int d^2 b \, \rho^{p, n} (b) = 1, 0$.
As emphasized in Ref.~\cite{Miller:2007uy}, they are proper densities 
and can be expressed as the overlap integrals of the light--cone wave 
functions with the same momentum and particle number. More generally,
they correspond to the $x$--integral of the impact parameter--dependent 
valence quark densities in the nucleon, which are defined as the
Fourier transform of the diagonal GPDs and describe the densities
of quarks minus antiquarks with respect to longitudinal momentum
fraction $x$ and transverse position $b$ \cite{Burkardt:2000za}
(see Fig.~\ref{fig:density}a).
Extensive numerical studies of the transverse charge densities 
have been performed using empirical parametrizations of the proton 
and neutron form factor data at spacelike momentum transfers; 
see Ref.~\cite{Venkat:2010by} for a recent summary and analysis 
of the uncertainties.

The nucleon form factors are analytic functions of the invariant 
momentum transfer $t$, with singularities (branch cuts, poles) on
the positive real axis. Assuming an asymptotic power behavior 
as $F_1 (t) \sim |t|^{-2}$, as expected from perturbative QCD
(with logarithmic modifications) and consistent with present
experimental data, the Dirac form factor satisfies a dispersion 
relation
\beq 
F_1^{p, n} (t) \;\; = \;\; 
\int\limits_{4m_\pi^2}^\infty \frac{dt'}{t' - t - i0} 
\; \frac{\textrm{Im}\, F_1^{p, n} (t')}{\pi} .
\label{dispersion}
\eeq
It expresses the form factor in terms of its imaginary part on the 
principal cut in the physical sheet at $t > 4 m_\pi^2$, also referred to as 
the spectral function. Physically, the singularities in the form
factor at $t > 0$ correspond to the transition of a timelike 
virtual photon to a hadronic state coupling to a nucleon--antinucleon
($N\bar N$) pair (see Fig.~\ref{fig:density}b). Most of the states of 
interest, such as the vector mesons $\rho$ and $\omega$ and their first 
excitations, lie below the $N\bar N$ threshold 
$t = 4 m_N^2 = 3.5 \, \textrm{GeV}^2$,
where the spectral functions cannot be measured directly in conversion 
experiments. However, theoretical methods can be used to constrain
the spectral functions in the unphysical region; details
will be given in Secs.~\ref{sec:isovector} and \ref{sec:isoscalar} below. 
Supplemented with such information and additional assumptions about 
the asymptotic behavior, the dispersion relations Eq.~(\ref{dispersion}) 
have been used to fit nucleon form factor data in the spacelike region 
and extract information about the spectral functions
\cite{Hohler:1976ax,Belushkin:2006qa}.

A new perspective on nucleon structure can be gained by combining the 
dispersion representation of form factors with the concept of transverse 
charge densities. Substituting Eq.~(\ref{dispersion}) in 
Eq.~(\ref{rho_fourier}) and carrying out the Fourier integral, one 
obtains the transverse charge density as a 
dispersion integral over the imaginary part of the Dirac form factor
in the timelike region \cite{Strikman:2010pu} 
\beq
\rho^{p, n} (b) \;\; = \;\; \int\limits_{4m_\pi^2}^\infty \frac{dt}{2\pi} 
\; K_0(\sqrt{t} b) \; \frac{\textrm{Im}\, F_1^{p, n} (t + i0)}{\pi} .
\label{rho_dispersion}
\eeq
This spectral representation of the transverse density has several 
interesting properties. First, thanks to the exponential drop--off 
of the modified Bessel function $K_0$ at large arguments,
\beq
K_0 (\sqrt{t} b) \;\; \sim \;\; [\pi / (2 \sqrt{t} b)]^{1/2}
\; e^{-\sqrt{t} b} 
\hspace{2em} (\sqrt{t} b \; \gg \; 1),
\label{K0_asymptotic}
\eeq
the dispersion integral converges exponentially at large $t$, in contrast
to the power--like convergence of the integral for the form 
factor, Eq.(\ref{dispersion}).
This greatly reduces the sensitivity to the high--energy region where 
the spectral function is poorly known. As an aside, we note that use of 
a subtracted dispersion relation in Eq.~(\ref{rho_fourier}) would lead 
to an expression for $\rho(b)$ which differs 
from Eq.~(\ref{rho_dispersion}) only by a term 
$\propto \delta^{(2)}(\bm{b})$; subtractions therefore have no influence 
on the dispersion result for the transverse density at finite $b$.
In this sense the representation Eq.~(\ref{rho_dispersion}) is similar
to the Borel transform used to eliminate polynomial terms in QCD
sum rules \cite{Shifman:1978bx}.

\begin{figure}
\includegraphics[width=0.31\textwidth]{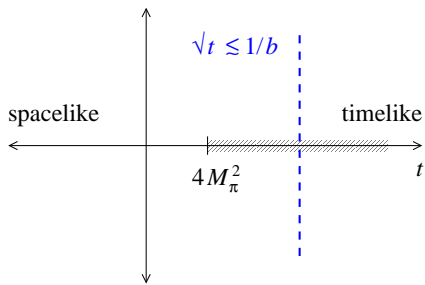}
\caption{The ``filtering'' property of the spectral representation 
of the transverse charge density, Eq.~(\ref{rho_dispersion}).
The dispersion integral extends over the cut of the form factor 
in the timelike region, $t > 4 m_\pi^2$. Because of the weighting
with $K_0 (\sqrt{t} b)$ only energies of the order 
$\sqrt{t} \lesssim 1/b$ in the spectral function are effectively sampled 
in the integral.}
\label{fig:tplane}
\end{figure}

Second, the transverse distance provides an external parameter which 
allows one to ``filter out'' a certain energy region in the 
spectral function. Because of the weighting with the kernel 
$K_0 (\sqrt{t} b)$ in Eq.~(\ref{rho_dispersion}) the dominant contribution 
to the integral for a given $b$ comes from energies in a range 
$\sqrt{t} \lesssim 1/b$ (see Fig.~\ref{fig:tplane}). 
This statement is to be understood in the sense of an exponential filter:
significant numerical suppression happens already at energies 
inside the range $\sqrt{t} \lesssim 1/b$, determining the overall magnitude 
of the resulting density; the important point is in the \textit{relative} 
suppression of higher energies (see Ref.~\cite{Miller:2010tz} for 
a detailed discussion). We shall use this property
in the following to identify regions in $b$ that are maximally 
sensitive to certain spectral regions of physical interest,
such as the near--threshold region 
$t - 4 m_\pi^2 \sim \textrm{few}\; m_\pi^2$ 
and the vector meson region $t \sim m_{\rho, \omega}^2$. The effectiveness 
of this method depends, of course, on the actual distribution of strength 
in the spectral functions and will be studied by numerical analysis.
More generally, this property will allow us to associate the nucleon's
partonic structure in the transverse periphery with the well--known
exchange mechanisms in the nucleon form factors.
%
%

Third, the dispersion representation is the proper mathematical
framework for discussing the asymptotic behavior of the transverse
densities in the limit of large $b$ and assess the uncertainties
of the empirical densities in the region where they are exponentially small. 
It is well--known that the asymptotic behavior of the Fourier transform 
[or, for that matter, the Fourier--Bessel transform 
Eq.~(\ref{rho_fourier_radial})] of a real function is 
determined by the singularities of that function in the complex plane,
as can be shown by deformation of the integration contour of the 
Fourier integral. Parametrizations of the spacelike 
form factors in terms of rational functions of $Q^2 = -t$ 
\cite{Friedrich:2003iz,Kelly:2004hm} generally
have unphysical singularities in the complex plane (e.g., pairs of
complex conjugate poles with finite imaginary part) that lead to 
a qualitatively wrong asymptotic behavior of the Fourier integrals
of Eqs.~(\ref{rho_fourier}) and (\ref{rho_fourier_radial}).
The only way to ensure qualitatively correct asymptotic behavior 
of the charge density is to use form factor parametrizations with the
proper analyticity, as provided by the dispersion 
representation of Eq.~(\ref{dispersion}). In this case the Fourier 
integral over spacelike $t$ becomes equivalent to the dispersion 
integral over timelike $t$, Eq.~(\ref{rho_dispersion}), and may be 
evaluated directly in this way. With the proper asymptotic form
ensured by the correct position of the singularities, one may then 
estimate the numerical uncertainty of the large--$b$ densities
from the uncertainty of the spectral strength at 
low $t$ \footnote{It is possible that some rational form factor fits
may give accurate numerical representations of the transverse 
density up to rather large values of $b$ when used in the Fourier
integral Eqs.~(\ref{rho_fourier}) and (\ref{rho_fourier_radial}).
However, the range of $b$ for which a given parametrization works 
depends on the exact location of the singularities in the complex plane, 
which is determined by the values of the fit parameters and may 
vary drastically between different fits. One therefore cannot advocate
this approach as a general method for studying the densities at large
$b$.}.

For theoretical analysis it is convenient to consider the isovector 
and isoscalar combinations of form factors and the corresponding 
transverse charge densities \footnote{In Ref.~\cite{Strikman:2010pu} we 
considered the difference of proton and neutron form factors without 
a factor $1/2$. In the present article we follow the standard convention 
for the isovector and isoscalar form factors with the factor $1/2$.}
\be
F_1^{V, S}(t) &\equiv& {\textstyle\frac{1}{2}} [F_1^p(t) \mp F_1^n(t)], 
\\
\rho^{V, S}(b) &\equiv& {\textstyle\frac{1}{2}} [\rho^p(b) \mp \rho^n(b)] ,
\label{rho_V_S}
\ee
which are normalized such that 
\beq
F^{V, S}(0) \;\; = \;\; \! \int d^2 b \, \rho^{V, S}(b) \;\; = \;\; 1/2 .
\eeq
Because they involve $t$--channel states of isospin 1 and 0, 
respectively, the two combinations have very different spectral functions.
In the following we discuss the spectral analysis of the transverse
charge densities separately for the isovector and isoscalar channels,
returning to the proton and neutron densities in Sec.~\ref{sec:pn}.

\section{Isovector charge density}
\label{sec:isovector}
\begin{figure}
\includegraphics[width=0.48\textwidth]{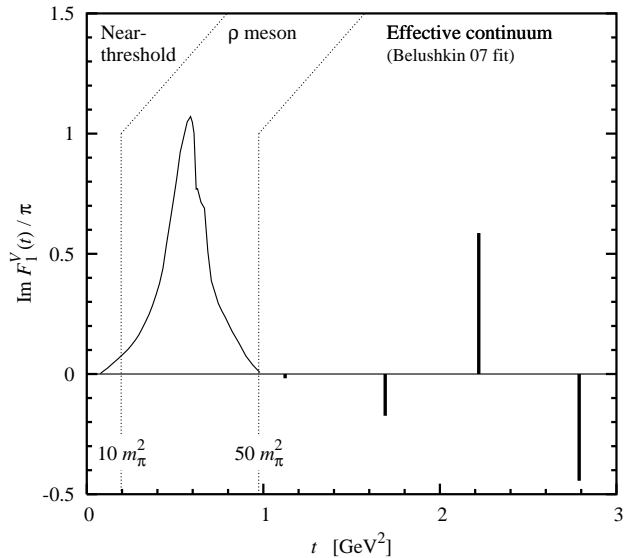}
\caption{Spectral function of the isovector nucleon Dirac form factor
$\textrm{Im} \, F_1^V (t)/\pi$ in the parametrization of 
Ref.~\cite{Belushkin:2006qa}. The dotted lines indicate the boundaries
of the three spectral regions discussed in the text: 
Near--threshold region $4\, m_\pi^2 < t < 10 \, m_\pi^2$; 
$\rho$ meson region, $10 \, m_\pi^2 < t < 50 \, m_\pi^2$; 
effective continuum region $t > 50 \, m_\pi^2$. The solid line shows 
the dispersion theory result for the $\pi\pi$ contribution in 
the near--threshold and $\rho$ meson regions. 
The spikes indicate the delta functions parametrizing the 
effective continuum; their absolute height is not drawn to scale,
but the relative heights reflect the ratio of coefficients determined 
in the SC fit \cite{Belushkin:2006qa}.}
\label{fig:spect_isovector}
\end{figure}

The spectral function of the isovector nucleon form factor has been
studied extensively in the literature and is under good theoretical 
control up to squared energies $t \sim 1 \, \textrm{GeV}^2$. Because the
isovector current couples to two pions, the threshold in this channel
is at $t = 4 \, m_\pi^2$. One can identify three distinct regions 
of the spectral function. At energies $t - 4 \, m_\pi^2 \sim 
\textrm{few} \, m_\pi^2$ the spectral function is governed by the 
universal threshold behavior implied by soft--pion dynamics and 
can be calculated in a model--independent manner. The traditional 
approach is to use dispersion theory to calculate the 
$\pi\pi \rightarrow N\bar N$ amplitude near threshold, taking care to
include the effect of a branch cut singularity on the unphysical sheet 
close to $t = 4 \, m_\pi^2$ \cite{hoehler:book,Belushkin:2005ds}.
Another approach is through chiral perturbation theory with 
relativistic nucleons, which naturally implements the correct analytic 
structure of the soft--pion amplitudes \cite{Gasser:1987rb,Kaiser:2003qp}.
At somewhat higher energies, $t \lesssim 50 \, m_\pi^2$, the spectral 
function is still saturated by the $\pi\pi$ channel, but rescattering 
effects play an important role away from threshold. In this region one 
can use elastic unitarity to calculate the spectral
function in terms of the measured $\pi\pi$ phase shifts, which are
dominated by the $\rho$ resonance \cite{hoehler:book,Belushkin:2005ds}. 
At even higher energies, $t \gtrsim 50 \, m_\pi^2$, the number of 
possible hadronic channels makes it impractical to calculate the 
spectral function from hadronic dynamics. However, it is constrained 
theoretically by the integral relations for the isovector charge 
(form factor at $t = 0$) and Dirac charge radius (derivative of the
form factor at $t = 0$) following from the dispersion integral
Eq.~(\ref{dispersion}), as well as the requirement that the
spacelike form factor drop faster than $|t|^{-1}$ at large momenta
(superconvergence relation). In this energy region one may use a series 
of poles as an effective parametrization of the hadronic
continuum, with the understanding that only their collective behavior,
not the individual masses and coefficients, carry physical significance.
Thus, in Ref.~\cite{Belushkin:2006qa} the isovector spectral function is
parametrized as
\be
\textrm{Im} \, F_1^V (t) &=& F_1^{V} (t)_{\pi\pi}  
\; + \; \sum_i^n a_i^V \delta(t - m_i^2) ,
\label{param_isovec}
\ee
where $F_1^{V} (t)_{\pi\pi}$ is the dispersion--theoretical result in the
$\pi\pi$ channel, covering the near--threshold and $\rho$ meson
region, and the poles parametrize the effective continuum;
the values of the parameters can be found in the quoted article.
In the superconvergence (SC) fit of Ref.~\cite{Belushkin:2006qa} 
the highest--mass singularity is actually parametrized as a broad 
resonance; this has practically no effect on our study of charge 
densities in the nucleon's periphery, as will be explained in
the following. Figure~\ref{fig:spect_isovector} shows the empirical 
spectral function in the three different regions. For illustration we
have chosen here $10 \, m_\pi^2$ as the upper boundary of the 
near--threshold region; alternative choices will be discussed below,
and our conclusions do not depend on the precise value.

%
%

\begin{figure*}
\begin{tabular}{ll}
\includegraphics[width=0.48\textwidth]{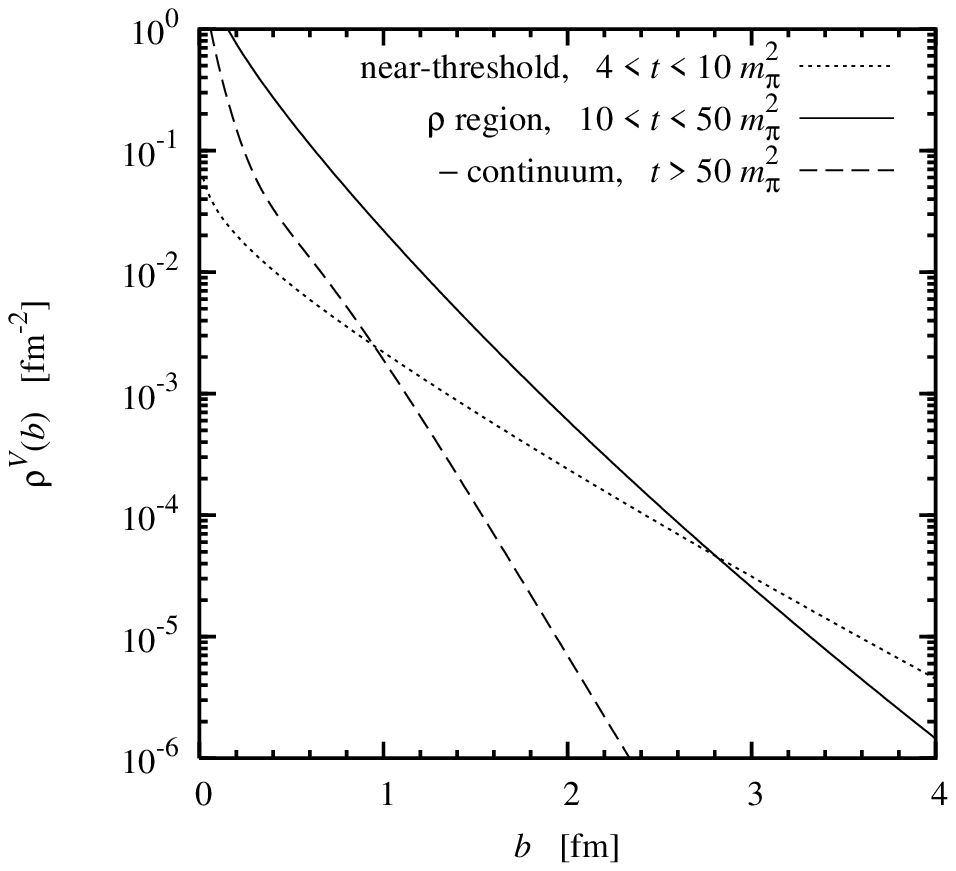}
&
\includegraphics[width=0.45\textwidth]{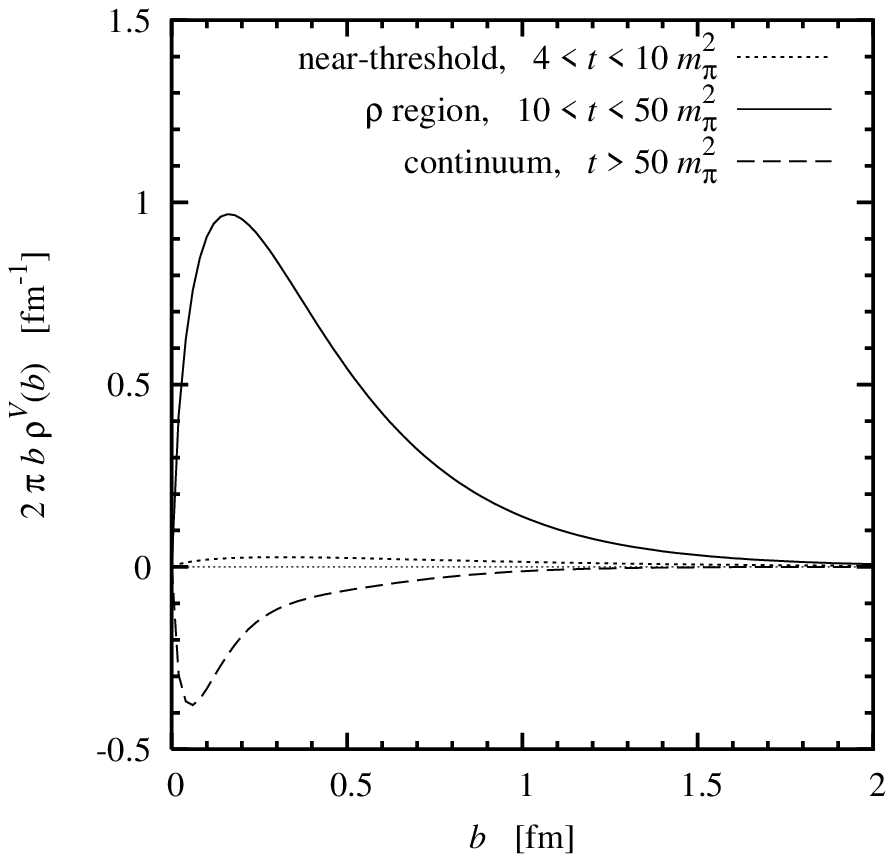}
\\[-3ex]
(a) & (b)
\end{tabular}
\caption{Contribution of different energy regions in the spectral 
function (cf.\ Fig.~\ref{fig:spect_isovector}) to the isovector 
charge density in the nucleon, $\rho^V (b)$,
calculated with the parametrization of Ref.~\cite{Belushkin:2006qa}
(SC fit). (a) Density $\rho^V (b)$ on a logarithmic scale.
(b) Radial density $2\pi b \rho^V (b)$ on a linear scale. 
Dotted lines: Near--threshold region $4 \, m_\pi^2 < t < 10 \, m_\pi^2$.
Solid lines: $\rho$ meson region $10 \, m_\pi^2 < t < 50 \, m_\pi^2$. 
Dashed lines: Effective continuum $t > 50 \, m_\pi^2$; this contribution
is negative and shown with reversed sign in the logarithmic plot (a).}
\label{fig:rhov}
\end{figure*}

Using this parametrization of the spectral function we can now quantify 
how much the different energy regions contribute to the isovector
charge density at a given $b$. The results are summarized in 
Fig.~\ref{fig:rhov}. Plot (a) shows the exponential fall--off of 
the various contributions to $\rho^V (b)$ on a logarithmic scale. 
Plot (b) shows the radial density $2\pi b \rho^V (b)$ on a linear scale; 
the integral of the total radial density, given by the area under the 
sum of the curves, is the total isovector charge, $1/2$. The results 
show several interesting features.
First, the near--threshold region $4 \, m_\pi^2 < t < 10 \, m_\pi^2$
is numerically important only at very large distances 
$b \gtrsim 2 \, \textrm{fm}$; see Fig.~\ref{fig:rhov}a. 
At smaller distances it is simply overwhelmed by the contribution of 
the $\rho$ meson region, which has a faster exponential decay but 
a much larger coefficient. This confirms the conclusion of 
Ref.~\cite{Strikman:2010pu}, that the chiral component in the 
nucleon's transverse charge density becomes clearly visible only 
at distances $b > 2 \, \textrm{fm}$. We note that the precise 
upper boundary of the near--threshold region is a matter of 
definition and depends on the requested accuracy of the
chiral expansion for the spectral function. For $t < 10\, m_\pi^2$ 
the leading--order chiral result accounts for more than half of the 
dispersion result (see Appendix~\ref{app:theoretical} and 
Fig.~\ref{fig:spect_f1}). However, it is not possible to 
substantially modify our conclusion by varying this value within 
reasonable bounds: a change from $t = 10 \, m_\pi^2$ to $15 \, m_\pi^2$ 
would give a near--threshold contribution to $\rho^V (b)$ that is 
1.7 times larger at $b = 2 \, \textrm{fm}$, which would have only a 
minor effect on the comparison with the $\rho$ region on the logarithmic 
scale of Fig.~\ref{fig:rhov}a. The important point here is that for 
any choice of boundary our approach allows us to quantify 
unambiguously how much the region thus defined contributes 
to the transverse density.
%
%

Second, over a wide range of intermediate distances 
$0.5 \lesssim b \lesssim 1.5\, \textrm{fm}$ the isovector transverse 
charge density is dominated by the $\rho$ meson mass region;
the high--mass continuum contribution reaches only $-12\%$ of the 
$\rho$ at $b = 0.5 \, \textrm{fm}$ and is substantially smaller
at larger $b$. The region $b \sim 1 \, \textrm{fm}$, where the
near--threshold contribution is equally small, represents the
cleanest case of  ``vector dominance'' in the transverse charge
density. Determination of the nucleon's transverse 
density in this region --- by reconstructing it from spacelike form 
factor data, or through theoretical calculations --- would thus provide 
unique information on the $\rho$ meson contribution to the spectral 
function and thus its coupling to the nucleon. Note that the 
dispersion representation Eq.~(\ref{rho_dispersion}) allows us 
to both maximize the sensitivity to the $\rho$ meson mass region 
and to quantify the corrections to vector dominance in a 
model--independent manner.

Third, the effective continuum contribution to the charge density 
remains relatively small down to distances as small as 
$b \sim 0.3 \, \textrm{fm}$. This happens because of the low spectral 
strength in the region immediately above the $\rho$, 
$1.0 < t < 1.4 \, \textrm{GeV}^2$, and the substantial cancellations 
between the higher--mass poles in the parametrization \cite{Belushkin:2006qa}
(see Fig.~\ref{fig:spect_isovector}). Whether the nucleon spectral
function in the region above the $\rho$ could at 
least partly be explained by the $\rho'$ resonances seen in the 
$\pi\pi$ channel is an interesting question which cannot 
be answered from form factor fits alone. We note that the 
$e^+e^- \rightarrow \pi\pi$ data clearly show a broad 
$\rho'$ resonance at $1.4\,\textrm{GeV}$ that interferes destructively
with the $\rho$ and results in a vanishing $\pi\pi$ strength at 
$t \sim 1.2 \, \textrm{GeV}^2$ (see Ref.~\cite{Bruch:2004py} and 
references therein), in qualitative agreement with the empirical strength 
distribution found in the nucleon form factor fit \cite{Belushkin:2006qa}.

In the SC fit of Ref.~\cite{Belushkin:2006qa} the highest--mass 
pole in Eq.~(\ref{param_isovec}) was actually replaced by a 
contribution to the form factor of the form
\beq
a_n^V (m_n^2 - t)/[(m_n^2 - t)^2 + \Gamma_n^2] ,
\label{broad}
\eeq
with $\Gamma_n$ comparable to $m_n^2$, 
mimicking the effect of a broad resonance.
As it stands, this term has poles in the complex plane away from the 
real axis, at $t = m_n^2 \pm i \Gamma_n$, and cannot be regarded as 
a contribution to the spectral function.
However, as can be seen by calculating the charge density from
the Fourier integral Eq.~(\ref{rho_fourier_radial})
over spacelike momentum transfers, the contribution of 
this term to the isovector density is very small at all but the 
shortest distances, $<2\%$ at 
$b > 0.1\, \textrm{fm}$ and $\ll 1\%$ at $b > 0.5\, \textrm{fm}$,
and we can safely neglect it in our study of the nucleon's periphery.
The same applies to the highest--mass pole in the isoscalar density 
considered in Sec.~\ref{sec:isoscalar}.

In Ref.~\cite{Strikman:2010pu} we studied the question at what
distances the isovector transverse charge density is dominated 
by chiral dynamics in a theoretical approach, by comparing the
chiral perturbation theory result for the transverse density at
$b \sim 1/m_\pi$ with the non--chiral density modeled 
by elementary $\rho$ meson exchange.
An interesting question is how the theoretical approach of
Ref.~\cite{Strikman:2010pu} relates to the present study of the 
transverse densities using empirical spectral functions.
This is explained in Appendix~\ref{app:theoretical},
where we summarize how well the empirical isovector spectral 
function in the different regions is reproduced by the theoretical
models used in Ref.~\cite{Strikman:2010pu}. Overall, the present
analysis with empirical spectral functions fully confirms our
earlier conclusion that the chiral component becomes numerically 
dominant only at distances $b \gtrsim 2 \, \textrm{fm}$,
contradicting naive expectations that the charge densities at
$b \gtrsim 1 \, \textrm{fm}$ could be attributed to the nucleon's 
``pion cloud.''
\section{Isoscalar charge density and its uncertainty}
\label{sec:isoscalar}
%
%
\begin{figure*}
\begin{tabular}{ll}
\includegraphics[width=0.48\textwidth]{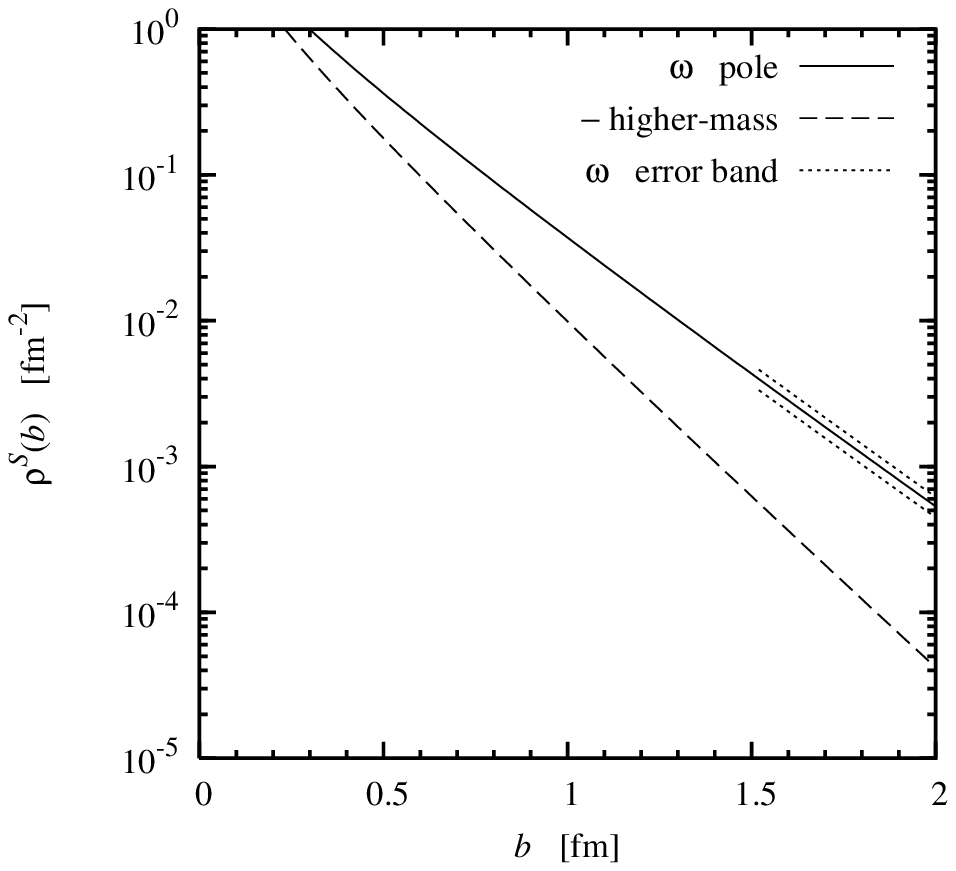}
&
\includegraphics[width=0.45\textwidth]{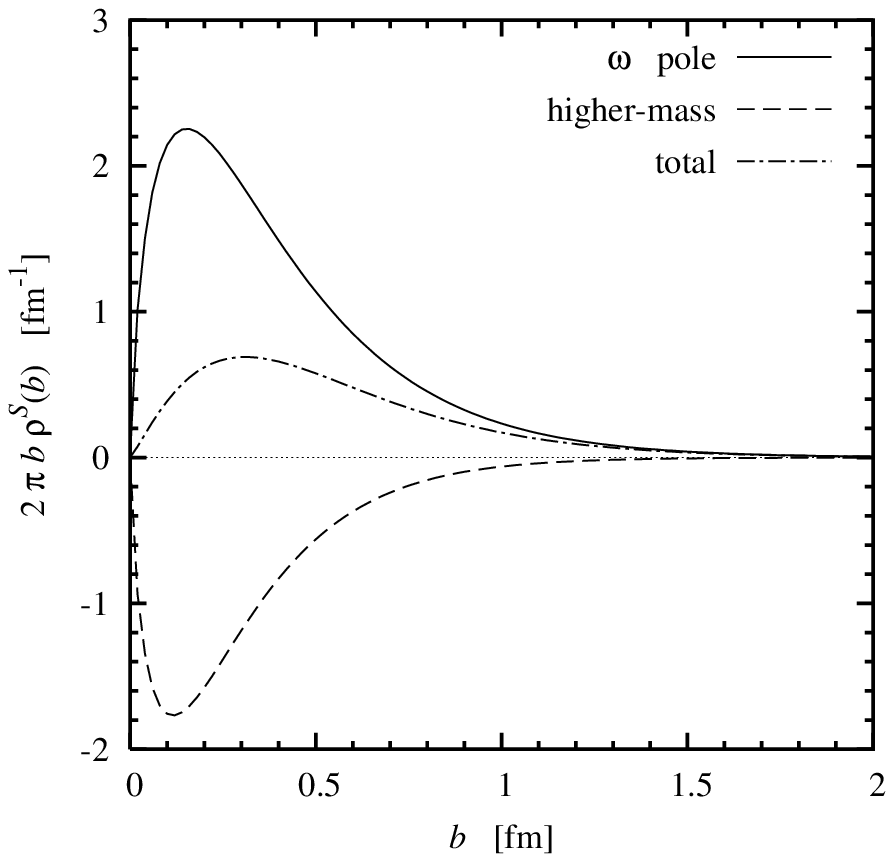}
\\[-3ex]
(a) & (b)
\end{tabular}
\caption{Contribution of different energy regions in the spectral 
function to the isoscalar charge density in the nucleon, $\rho^S (b)$,
calculated with the parametrization of Ref.~\cite{Belushkin:2006qa}
(SC fit). (a) Density $\rho^S (b)$ on a logarithmic scale.
Solid line: $\omega$ meson pole. Dashed line: Higher--mass states 
$t > 1\, \textrm{GeV}^2$, including both identified hadronic states 
($K\bar K, \rho\pi$) and the effective continuum (this contribution 
is negative and shown with reversed sign). Dotted lines: Error band 
of the $\omega$ contribution, giving approximately the uncertainty 
of the total isoscalar density at $b > 1.5 \, \textrm{fm}$.
(b) Radial density $2\pi b \rho^S (b)$ on a linear scale. 
Solid line: $\omega$ meson pole. Dashed line: Higher--mass 
states $t > 1\, \textrm{GeV}^2$. Dot--dashed line: Total.}
\label{fig:rhos}
\end{figure*}
The isoscalar spectral function at low energies behaves very differently 
from the isovector one, and comparatively little is known about it
from first principles. The lowest hadronic state in the isoscalar
channel allowed by quantum numbers is the $3\pi$ state. The non-resonant
$3\pi$ contribution near threshold was estimated using heavy--baryon 
chiral perturbation theory \cite{Bernard:1996cc} and found to be 
roughly two orders of magnitude smaller than the $2\pi$ contribution 
in the isovector channel; it therefore plays no role in the transverse
charge density at the distances $b \sim 2-3 \, \textrm{fm}$ of
interest here (cf.\ Fig.~\ref{fig:rhov}a). The strength in the $3\pi$ 
channel is overwhelmingly concentrated in the $\omega$ resonance at 
$m_\omega = 0.782\, \text{GeV}$, 
whose width can be neglected for our purposes. At energies 
$\sqrt{t} \gtrsim 1 \, \textrm{GeV}$ other hadronic channels 
come into play. The $K\bar K$ contribution was computed using
dispersion theory \cite{Hohler:1974ht,Hammer:1999uf} and exhibits 
the $\phi$ resonance at 1.02 GeV,
very close to threshold; in contrast to $\pi\pi$ in the isovector 
channel there is no enhancement of the strength to the left of
the resonance. In the parametrization of Ref.~\cite{Belushkin:2006qa}
the entire $K\bar K$ strength is described by an effective pole
at the $\phi$ mass. Additional strength in this region is expected
to come from the $\pi\rho$ continuum, which was found to be sizable
in the context of the Bonn--J\"ulich meson exchange model 
of the $NN$ interactions \cite{Meissner:1997qt}. This contribution
is again parametrized by an effective pole. We emphasize that the details of
the theoretical estimates of these explicit higher--mass contributions 
are ultimately not essential for the accuracy of the parametrization
of the spectral function in Ref.~\cite{Belushkin:2006qa}, as these states 
have masses of the same order as the effective continuum poles,
whose strength is determined by the fit to the form factor data.

For the purpose of our analysis, we divide the empirical isoscalar 
spectral function into the $\omega$ pole, which is the analogue of 
the $\rho$ in the isovector channel and accounts for the entire strength 
at energies $\sqrt{t} < 1 \, \textrm{GeV}$, and a ``rest'' of
higher--mass states, about whose nature we remain agnostic 
at this point. The respective contributions to the isoscalar transverse
charge density are shown in Fig.~\ref{fig:rhos}.
One sees that the relative contribution from higher--mass states 
is substantially larger than in the isovector density, amounting
to $-27 \%$ of the $\omega$ at $b = 1\, \textrm{fm}$. Vector dominance
at intermediate distances is therefore realized not as perfectly as 
in the isovector charge density. However, because of the
absence of a non-resonant contribution below the $\omega$ mass,
in the isoscalar case one has the option to go to larger distances 
to maximize the vector meson contribution: at $b = 2 \, \textrm{fm}$
the contribution from higher--mass states has dropped to $-8 \%$
of the $\omega$. Thus, it is possible to realize ``vector meson dominance''
in the isoscalar charge density as well.

In view of the paucity of theoretical information in the isoscalar
sector, it is worthwhile to consider the uncertainty of the empirical 
isoscalar transverse density at large $b$. In the region where it is 
dominated by the $\omega$ contribution its uncertainty is essentially 
determined by the accuracy with which the coefficient of the
$\omega$ pole can be determined from dispersion fits to the isoscalar 
form factor. The analysis of Ref.~\cite{Belushkin:2006qa} quotes an 
uncertainty of $\pm 16\%$ for the $\omega$ coefficient; 
we can therefore ascribe a relative uncertainty of this magnitude to 
the isoscalar charge density at $b > 1.5 \, \textrm{fm}$, where the
$\omega$ accounts for more than $80\%$ of the total density
(see Fig.~\ref{fig:rhos}). (An even larger range of $\omega NN$ 
couplings is quoted in Ref.~\cite{Diehl:2007uc}; however, that 
analysis uses a more restrictive form factor fit than the one of 
Ref.~\cite{Belushkin:2006qa}.) At smaller values of $b$ the contribution
from higher--mass poles can no longer be neglected and correlations
between the errors of the coefficients of the various poles
become important in estimating the error of the total charge density;
unfortunately, this information is not provided in the fit
of Ref.~\cite{Belushkin:2006qa}. Altogether, we see that there is
considerable uncertainty in the isoscalar charge density at large $b$.

%
%
\begin{figure}
\includegraphics[width=0.48\textwidth]{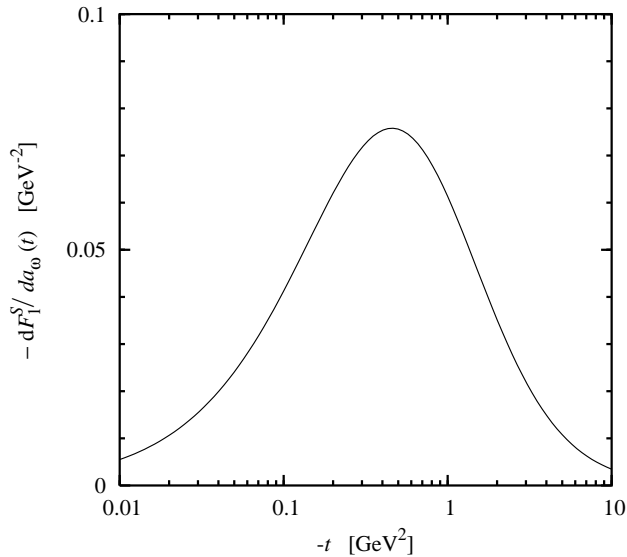}
\caption{Sensitivity of the pole fit to the isoscalar form factor,
Eq.~(\ref{pole_fit}), to the $\omega$ coefficient $a_\omega$.
The curve shows the derivative of the form factor with respect
to $a_\omega$ after the constraints Eqs.~(\ref{constraint1})
and (\ref{constraint2}) were used to eliminate two of the 
other parameters. Here $N = 3$, and the pole masses are
those of the SC fit of Ref.~\cite{Belushkin:2006qa}.}
\label{fig:fit}
\end{figure}
In order to determine more accurately 
the isoscalar transverse density in the nucleon's 
periphery it is obviously necessary to gain better 
knowledge of the coefficient of the $\omega$ pole in the isoscalar 
spectral function (or, equivalently, the $\omega NN$ coupling) 
from dispersion fits to spacelike form factor data. It is interesting 
to ask at what momentum transfers future form factor measurements 
would have the strongest impact on the determination of the $\omega$ 
coefficient. Naively one might think that, because the $\omega$ 
completely dominates the charge density at $b \gtrsim 2 \, \textrm{fm}$, 
form factor data at $-t 
\lesssim \pi^2 / (2 \, \textrm{fm})^2 = 0.1 \, \textrm{GeV}^2$
would be most useful to constrain the $\omega$ coefficient. However, 
it is data over a broad range of intermediate momentum transfers 
$-t \sim m_\omega^2$
that effectively determine the $\omega$ coefficient in the dispersion
analysis. The reason lies in the nature of the spectral representation ---
the $\omega$ is the leading singularity, and all spacelike momenta
in the range $-t \sim m_\omega^2$ are equally affected by the strength 
of this pole. To see this explicitly, let us consider a spectral
representation of the isoscalar form factor as an $\omega$ pole and 
a sum of $n - 1$ higher--mass poles \cite{Hohler:1976ax,Belushkin:2006qa}
\be
F_1^S (t) &=& \frac{a_\omega}{m_\omega^2 - t}
\; + \; \sum_{i = 2}^n \frac{a^S_i}{m_i^2 - t} .
\label{pole_fit}
\ee
The coefficients are constrained by charge conservation and the
$|t| \rightarrow \infty$ asymptotic behavior of the form factor,
\be
F_1^S (0) &=& 1/2, 
\label{constraint1}
\\[1ex]
\lim_{|t| \rightarrow \infty} \; t \, F_1^S (t) &=& 0 . 
\label{constraint2}
\ee
The resulting linear relations allow one to express two coefficients
in terms of the other $n - 2$. A value $n \geq 3$ is required
to have sufficient flexibility in the fit and avoid artificial
correlations between the behavior at small and large $|t|$.
The fit of Ref.~\cite{Belushkin:2006qa} effectively works with
$n = 4$ \footnote{In the fit of Ref.~\cite{Belushkin:2006qa}
the calculated $\rho\pi$ continuum is approximated by a pole 
with a mass of $1.125 \, \textrm{GeV}$, practically identical to
the pole representing the calculated $K\bar K$ continuum and the
explicit $\phi$ contribution, such that all these contributions
effectively amount to a single pole at the $\phi$ mass.};
that its highest--mass pole has a finite width, 
Eq.~(\ref{broad}), is not important for our argument here. 
Figure~\ref{fig:fit} shows the derivative of the form factor 
parametrization Eq.~(\ref{pole_fit}) with respect to $a_\omega$ 
after the constraints Eqs.~(\ref{constraint1}) and (\ref{constraint2}) 
were used to eliminate two of the other coefficients, for $n = 4$
and the mass values of Ref.~\cite{Belushkin:2006qa}
(the $a_\omega$--derivative does not depend on the value of the 
remaining free coefficient but only on the position of the poles).
The result clearly shows that the sensitivity to $a_\omega$ is
broadly distributed over a range of momentum transfers 
$|t| \sim m_\omega^2$, suggesting that precise form factor measurements
in this region would be most useful to constrain this parameter.
A more accurate analysis of the impact of future form factor data
on the determination of the large--$b$ isoscalar densities,
with account of experimental uncertainties and correlations between 
parameters, remains an interesting problem for further study.
\section{Proton and neutron charge densities}
\label{sec:pn}
\begin{figure}
\includegraphics[width=0.48\textwidth]{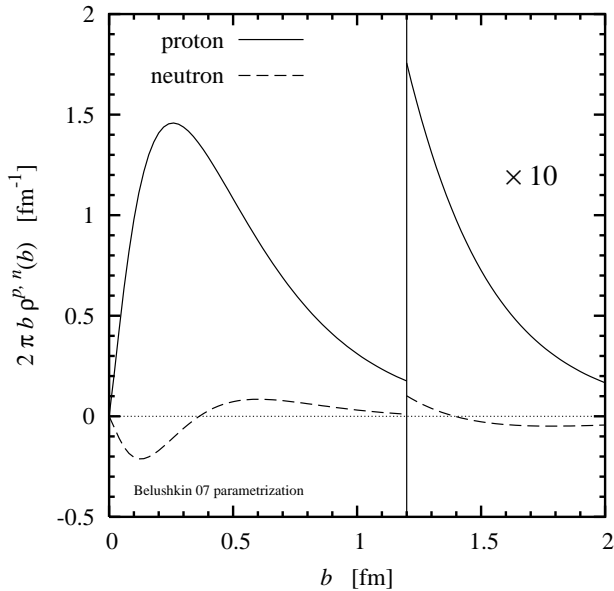}
\caption{Transverse charge densities in the proton (solid line)
and neutron (dashed line) obtained from the dispersion integral
Eq.~(\ref{rho_dispersion}) with the empirical spectral functions
of Ref.~\cite{Belushkin:2006qa}.}
\label{fig:rhopn}
\end{figure}

Using the dispersion results for the isovector and isoscalar transverse 
densities we can construct the transverse charge densities in the 
proton and neutron, cf.\ Eq.~(\ref{rho_V_S}). The results are
shown in Fig.~\ref{fig:rhopn}. The dispersion integral 
Eq.~(\ref{rho_dispersion}) with the spectral functions of
Ref.~\cite{Belushkin:2006qa} gives a peripheral charge density in 
the neutron that is clearly negative above 
$b \gtrsim 1.5\, \textrm{fm}$, and positive over a wide range 
of intermediate distances $b \sim 0.5 - 1.5 \, \textrm{fm}$.
A positive density at such distances was found previously 
in a Fourier analysis of the spacelike neutron form 
factor \cite{Miller:2007uy}. With the insights into the spectral
composition of the transverse change densities from the studies
of Secs.~\ref{sec:isovector} and \ref{sec:isoscalar} we can now
explain this behavior of the neutron charge density from 
the $t$--channel point of view.
%
%

The negative charge density in the neutron at large distances 
arises because the spectral strength in the isovector channel
starts at lower masses than in the isoscalar channel, namely 
$4 m_\pi^2$ rather than $m_\omega^2$. As a result, the isovector
density has a slower exponential decay and becomes dominant
in the $b \rightarrow \infty$ limit (note that $\rho^n = \rho^S - \rho^V$). 
This is a robust prediction of the dispersion approach, which is 
independent of the details of the parametrization of the spectral 
functions. Qualitatively, such large--distance behavior is consistent 
with the the picture of the neutron as a proton at the center and a 
negative pion in the cloud. However, the analysis of Sec.~\ref{sec:isovector}
shows that the chiral near--threshold region of the isovector spectral 
function becomes numerically dominant only at very large distances 
$\gtrsim 4 \, \textrm{fm}$ (see Fig.~\ref{fig:rhov}a). At the distances 
of interest here, $b \sim 2 \, \textrm{fm}$, the isovector density 
results rather from the broadly distributed strength in 
the $\rho$ meson region. We conclude that non--chiral interactions
still play an essential role in the transverse density at such 
distances. That the ``pion cloud'' is not yet dominant at 
$b \sim 2 \, \textrm{fm}$ is also seen from the fact that the
proton and neutron densities are still far from being equal
and opposite in sign, because of the large isoscalar density
arising from the $\omega$.

\begin{figure}
\includegraphics[width=0.48\textwidth]{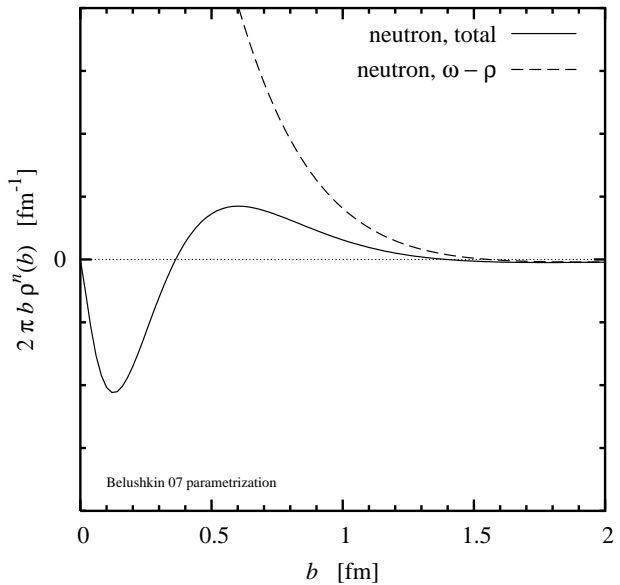}
\caption{Spectral composition of the transverse charge density 
in the neutron, as evaluated with the spectral functions
of Ref.~\cite{Belushkin:2006qa}. Solid line: Total dispersion result
(cf.\ Fig.~\ref{fig:rhopn}).
Dashed line: Contribution from $\sqrt{t} < 1 \, \textrm{GeV}$,
given by the difference $\omega - \rho$ (here $\rho$ includes
the near--threshold region).}
\label{fig:neutron}
\end{figure}

The positive density in the neutron at intermediate distances
$b \sim 1 \, \textrm{fm}$ lies in the region where vector 
mesons give a prominent contribution to the isovector and 
isoscalar transverse densities; cf.\ Secs.~\ref{sec:isovector}
and \ref{sec:isoscalar}. An interesting question is whether 
the positive charge density in the neutron could be explained 
solely on the basis of the vector meson region in the 
spectral functions, i.e., as the result of vector meson exchange 
in the form factor. To answer this question one needs to look 
in detail at the spectral composition of the neutron charge density
in the region $b \sim 0.5 - 1.5 \, \textrm{fm}$. Figure~\ref{fig:neutron}
shows the total neutron charge density obtained from the dispersion
integral, as well as the result from the region 
$\sqrt{t} < 1 \, \textrm{GeV}$, corresponding to the difference 
of the $\omega$ and $\rho$ region of the isoscalar and isovector 
spectral functions, respectively (here the near--threshold region
is included in the $\rho$; but its contribution is numerically small, 
see Fig.~\ref{fig:rhov}). One sees that the vector meson region
alone does produce a positive neutron charge density; however, 
with the $\omega NN$ coupling of Ref.~\cite{Belushkin:2006qa}, this
contribution is substantially larger than the full dispersion result. 
Higher--mass states, particularly in the isoscalar channel, are essential 
for explaining the positive neutron charge density at the quantitative level.
%
%

In sum, our $t$--channel analysis shows that the precise value of 
the transverse charge density in the neutron at distances 
$b \sim 0.5 - 1.5 \, \textrm{fm}$ is closely tied up with the question 
of the spectral strength in the isoscalar channel at masses 
$\sqrt{t} \sim 1\, \textrm{GeV}$. As shown in 
Sec.~\ref{sec:isoscalar}, the information on the transverse density
in this region comes from form factor measurements over a broad range of
intermediate momentum transfers $|t| \sim 0.1 - 1\, \textrm{GeV}^2$
(see Fig.~\ref{fig:fit} for the isoscalar component). Accurate
measurements of the neutron form factor at these momentum transfers
may thus considerably improve our knowledge of the isoscalar spectral
function. Because of the potential contribution from mesons containing
strange quarks ($K\bar K, \phi$), this question is of interest also for 
the determination of the strangeness content of the nucleon;
see Ref.~\cite{Diehl:2007uc} and references therein.

The transverse charge density in the neutron at large $b$ was recently 
studied by evaluating the Fourier transform of the spacelike form factor 
Eq.~(\ref{rho_fourier_radial}) \cite{Vanderhaeghen:2010nd}, using an 
updated version of the Friedrich--Walcher form factor 
parametrization \cite{Friedrich:2003iz} that includes recent
data from the BLAST \cite{Geis:2008ha} and Jefferson Lab 
Hall A experiments \cite{Riordan:2010id}. For $b < 2\, \textrm{fm}$
their Fourier result agrees well with the neutron density obtained from 
the dispersion integral Eq.~(\ref{rho_dispersion}) with the spectral 
functions of Ref.~\cite{Belushkin:2006qa},
with a maximum discrepancy of $\sim 20\%$ at $b = 1.7 \, \textrm{fm}$.
At $b > 2.4\, \textrm{fm}$ the Fourier result 
of Ref.~\cite{Vanderhaeghen:2010nd} becomes positive,
in contradiction to the robust prediction of the dispersion approach
(see above). This behavior of the Fourier transform may be a consequence 
of the fact that the spacelike form factor fit of Ref.~\cite{Friedrich:2003iz}
uses a higher--order rational function with unphysical singularities 
in the complex $t$--plane; cf.\ the discussion in Sec.~\ref{sec:spectral}.
\section{Implications for partonic structure}
\label{sec:partonic}
The transverse densities obtained from the dispersion representation
of the nucleon form factors provide interesting insight into the
nucleon's partonic structure. Here we would like to point out several 
implications that can be stated in a model--independent manner.

\begin{figure}
\includegraphics[width=0.48\textwidth]{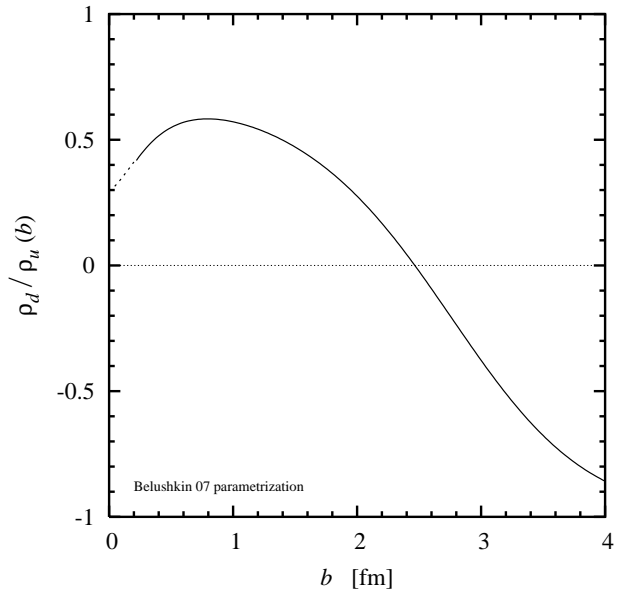}
\caption{Ratio of $d$-- and $u$--quark transverse valence
quark densities in the proton, $\rho_d (b) / \rho_u(b)$, 
as a function of $b$, as obtained from the dispersion integral
Eq.~(\ref{rho_dispersion}) evaluated with the spectral functions
of Ref.~\cite{Belushkin:2006qa}.}
\label{fig:durat}
\end{figure}

For the partonic interpretation of our results it is convenient to 
extract the transverse densities of $u$ and $d$ valence quarks in 
the proton, defined as the integral over $x$ of the impact 
parameter--dependent valence quark densities:
\beq
\rho_u (b) \;\; \equiv \;\; \int_0^1 \! dx \; [u(x, b) - \bar u(x, b)], 
\hspace{2em} \text{etc.}
\eeq
They are related to the isoscalar and isovector charge densities by
\beq
\rho_{u, d}(b) \;\; = \;\; 3\rho^S(b) \pm \rho^V(b)
\eeq
and normalized such that $\int d^2 b \, \rho_{u, d}(b) = 2, 1$.
Figure~\ref{fig:durat} shows the ratio of $d$-- and $u$--quark 
transverse valence quark densities, $\rho_d (b) / \rho_u(b)$, 
as obtained from the dispersion integral Eq.~(\ref{rho_dispersion}) 
evaluated with the spectral functions of Ref.~\cite{Belushkin:2006qa}.
The numerical result exhibit several interesting features.
In the limit of large transverse distances we expect that 
\beq
\rho_d (b) / \rho_u(b) \;\; \rightarrow \;\; -1 
\hspace{2em} (b \rightarrow \infty).
\eeq
In the $t$--channel (or exchange mechanism) view this should
happen because at asymptotically large $b$ the isovector charge 
density due to chiral two--pion exchange near threshold should
become dominant; see Secs.~\ref{sec:isovector} and \ref{app:theoretical}.
In the $s$--channel (or partonic) view the transverse density
at such distances should result from configurations in the proton's
light--cone wave function corresponding to a neutron at the center
and a peripheral $\pi^+$, which contribute to the $u$ and $\bar d$
densities in the proton. The numerical results show that the ratio
becomes negative at $b > 2.5 \, \textrm{fm}$ but is still far from
$-1$, reaffirming our earlier conclusion that the chiral component 
becomes numerically dominant only at substantially larger distances.
%
%

Over a broad range of intermediate transverse distances 
the $d/u$ valence quark ratio in Fig.~\ref{fig:durat} is
\beq
\rho_d (b) / \rho_u(b) \;\; \approx \;\; 1/2 
\hspace{2em}
(0.2 \, \textrm{fm} \lesssim b \lesssim 1.6 \, \textrm{fm}).
\eeq
This value would be obtained in a mean--field picture of the 
valence quark configurations in the nucleon, in which $u$ and $d$ 
quarks move approximately independently on identical orbitals and the 
ratio of densities is given just by that of the quark numbers.
We thus have model--independent evidence for an approximate
mean--field picture of the nucleon's valence quark structure
at non--exceptional distances. In the $t$--channel view this
region is governed by the vector mesons, albeit with a 
non--negligible contribution from higher--mass states in the
isoscalar channel. Exploring this duality between valence quark
structure and vector meson exchange in dynamical models of 
the nucleon would be an interesting problem for further study.

The same conclusion could in principle be reached already 
from inspection of the proton and neutron densities in 
Fig.~\ref{fig:rhopn}. The neutron density in the region 
$0.3 < b < 1.5 \, \textrm{fm}$ is substantially smaller than the
proton one; in a picture of independent particles the former 
would be zero. The advantage of using the ratio $\rho_d /\rho_u (b)$ 
is that it eliminates much of the non--trivial $b$--dependence 
on the bound--state structure in this region.

Finally, at even smaller distances, $b < 0.2 \, \textrm{fm}$, the 
charge density ratio obtained from the dispersion representation
of Ref.~\cite{Belushkin:2006qa} drops significantly below 
the mean--field value of $1/2$ (see Fig.~\ref{fig:durat})
\footnote{At $b < 0.2\, \textrm{fm}$ the densities shown in
Fig.~\ref{fig:durat} (dotted line) were calculated as the Fourier 
integral of the spacelike form factor parametrization of 
Ref.~\cite{Belushkin:2006qa}, in order to include the contribution 
of the highest--mass pole of the form Eq.~(\ref{broad}).}.
At such values of $b$ it is 
possible that part of the proton charge density results 
from partonic configurations in which the active quark carries large 
momentum fraction $x \sim 1$, while the spectators are
restricted to substantially smaller values \cite{Miller:2008jc}. 
(The variable $b$ measures the distance of the active
parton from the transverse center--of--mass of the nucleon, 
see Fig.~\ref{fig:density}a \cite{Burkardt:2000za}. 
For $x \rightarrow 1$ the center--of--mass coincides with 
the position of the active parton, whence such configurations contribute
to the density at $b \rightarrow 0$ independently of their
physical transverse size; see Ref.~\cite{Miller:2010tz} 
for a detailed discussion.) The observed behavior of $\rho_d (b)/\rho_u (b)$
is consistent with a decreasing ratio of $d$ and $u$ valence
quark densities at large $x$,
\beq
[d(x) - \bar d(x)] / [u(x) - \bar u(x)] \; \ll \; 1
\hspace{2em} (x \rightarrow 1).
\eeq
Experimental information on this ratio at large $x$ is dependent on 
theoretical corrections to nuclear binding effects in measurements 
with nuclear targets, which are the subject of on--going research;
new data are expected with the 12~GeV Upgrade of 
Jefferson Lab \cite{12GeV_CDR}.
A joint analysis of elastic form factors and large--$x$ inclusive 
scattering data could explore the properties of large--$x$ 
configurations in the nucleon's partonic wave function more 
effectively than either class of observables alone.

We note that at distances $b < 0.2\,\textrm{fm}$ the proton and neutron 
charge densities defined by Eqs.~(\ref{rho_fourier}) and 
(\ref{rho_fourier_radial}) are sensitive to the spacelike form
factors at high momentum transfers where experimental information 
is limited, especially for the neutron. In particular, the
$u$-- and $d$--quark densities at $b = 0$ can be obtained from
the ordinary (not Fourier) integrals of the spacelike form factors as
\beq
\rho_{u, d}(b = 0)  \;\; = \;\; 
\int_0^\infty \! \frac{dQ^2}{4\pi} \; F_{1u, d} (t = - Q^2) ,
\eeq
where $F_{1u} \equiv 2F_1^p + F_1^n$ and $F_{1d} \equiv 2F_1^n + F_1^p$
are the $u$-- and $d$--quark contributions to the form factor.
A recent analysis \cite{Cates:2011pz}
including the neutron data from the Jefferson Lab 
Hall A experiment \cite{Riordan:2010id} extracted the $u$-- and 
$d$--quark Dirac form factors up to $Q^2 = 3.4\, \textrm{GeV}^2$.
Integration of these data, assuming extrapolation into the 
unmeasured high--$Q^2$ region by a rational fit with a leading $1/Q^4$ 
behavior, gives a ratio $\rho_d(0)/\rho_u (0) \approx 0.35$, somewhat 
larger than the value $0.3$ obtained from Ref.~\cite{Belushkin:2006qa} 
(see Fig.~\ref{fig:durat}), but still substantially below $1/2$.
We note that at $b > 0.2 \, \textrm{fm}$ the charge densities obtained 
from these data are in good agreement with those obtained from the 
dispersion fit of Ref.~\cite{Belushkin:2006qa}.
With the 12~GeV Upgrade of Jefferson Lab the neutron's Dirac form
factor will be measured up to $Q^2\sim 8\, \textrm{GeV}^2$ \cite{12GeV_CDR},
substantially reducing the uncertainties in the $u$-- and $d$--quark
densities at small $b$.
\section{Summary and outlook}
\label{sec:summary}
The dispersion approach to transverse densities allows one
to formulate the concept of vector dominance in the nucleon
form factors in a manner which is fully quantitative and
consistent with QCD. Extraction of the transverse densities
in the region $b \sim 1\, \textrm{fm}$ can provide unique 
information on the $\rho$ meson's coupling to the nucleon.
It also affords a model--independent definition that could 
in principle serve as a basis for calculating this hadronic 
coupling using non--perturbative QCD methods, such as lattice 
calculations.

The spectral analysis of transverse densities at intermediate 
distances $b \sim 1 \, \textrm{fm}$ suggests an interesting 
connection between vector dominance and the valence quark structure
of the nucleon. Such duality might be realized in a 
relativistic constituent quark picture, where the leading 
singularity ``seen'' by the current is at a mass 
$\sqrt{t} = 2 m_{\rm const} \approx m_\rho$. An effective
dynamics of chiral constituent quarks at a low resolution scale
appears as a result of the spontaneous breaking of chiral symmetry 
in QCD. Exploring this connection in explicit dynamical models 
would be an interesting problem for further study.

Generally, the dispersion representation Eq.~(\ref{dispersion}) 
provides the proper mathematical framework for studying transverse 
densities at distances $b \gtrsim 1 \, \textrm{fm}$. Its analyticity ensures 
the correct asymptotic behavior of the density, and the exponential 
fall--off of the different contributions is encoded already in the 
position of the singularities. It thus represents a valuable tool 
for studying peripheral nucleon structure using empirical or
theoretical methods. Dispersion fits to the spacelike nucleon form 
factors therefore have a special significance and should be given 
high priority as a method of data analysis \cite{Belushkin:2006qa}.
Such fits should be updated as new data become available, 
particularly with the 12 GeV Upgrade of Jefferson Lab that will
cover the high--$Q^2$ region with high precision. One should 
also explore improved parametrizations of the spectral functions 
in the high--mass region that satisfy QCD constraints and respect 
the analytic properties of the form factor.

Neutron form factor data are of particular importance for extracting 
the $\omega NN$ coupling and, indirectly, the coupling of higher--mass 
states in the isoscalar channel possibly related to the nucleon's 
strangeness content. Our estimates show that these objectives 
require accurate measurements over a broad range of intermediate momentum
transfers $|t| \sim 0.1-1\,\textrm{GeV}^2$ rather than exceptionally 
large or small values.

A similar spectral analysis could be performed for the nucleon's 
Pauli form factor, whose partonic representation is related to the
angular momentum of partons in the light--cone wave function;
such analysis is in progress. The approach described here could
also be extended to the axial form factors, whose transverse representation
constrains the quark helicity distributions in the nucleon.
\section*{Acknowledgments}
G.~A.~M.\ acknowledges the hospitality of Jefferson Lab and the University
of Adelaide during the work on this study. This work is supported by the 
U.S.\ DOE under Grants No. DE-FGO2-97ER41014 and DE-FGO2-93ER40771.
The work of G.~A.~M.\ is also partially supported  by the Director, 
Office of Energy Research, Office of High Energy and Nuclear Physics, 
Divisions of Nuclear Physics, of the U.S.\ Department of Energy under 
Contract No.~DE-AC02-05CH11231
Notice: Authored by Jefferson Science Associates, LLC under U.S.\ DOE
Contract No.~DE-AC05-06OR23177. The U.S.\ Government retains a
non--exclusive, paid--up, irrevocable, world--wide license to publish 
or reproduce this manuscript for U.S.\ Government purposes.
\appendix
\section{Theoretical analysis of isovector charge density}
\label{app:theoretical}
In Ref.~\cite{Strikman:2010pu} we studied the question at what
transverse distances the isovector charge density in the nucleon 
is dominated by the universal chiral dynamics that governs the
long--range behavior of strong interactions. The large--$b$ limit 
of the isovector charge density is determined by the threshold behavior 
of the spectral function near $t \rightarrow 4 \, m_\pi^2$, corresponding
to $t$--channel exchange of two soft pions, which can be analyzed
in a model--independent manner within chiral perturbation 
theory \cite{Gasser:1987rb,Kaiser:2003qp}.
By comparing the calculated chiral contribution to the non--chiral density 
arising from zero--width $\rho$ meson exchange we found that the former
becomes numerically dominant only at distances $b \gtrsim 2 \, \textrm{fm}$.
In this appendix we explain how the theoretical approach of 
Ref.~\cite{Strikman:2010pu} relates to the present dispersion analysis 
of the transverse densities, by showing how well, and in what sense, 
the theoretical approximations used in Ref.~\cite{Strikman:2010pu} reproduce 
the empirical spectral functions. This also allows us to address 
some questions concerning the quantitative comparison of ``chiral'' and
``non--chiral'' components of the transverse density that were not
discussed in detail in Ref.~\cite{Strikman:2010pu}, such as the role 
of higher--order chiral corrections, the finite width of the $\rho$, 
and the value of the $\rho NN$ coupling. 

\begin{figure}
\includegraphics[width=0.48\textwidth]{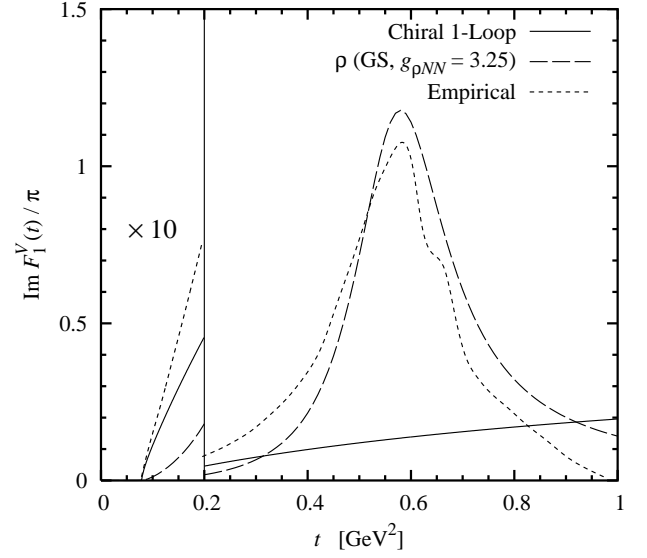}
\caption{Comparison of theoretical approximations to the isovector 
spectral function  with the empirical result of Ref.~\cite{Belushkin:2005ds}.
Solid line: Chiral one--loop result, Eq.~(\ref{imag}). Dashed line: 
$\rho$ meson contribution [GS form Eq.~(\ref{isovec_gs}), 
$\Gamma_\rho = 0.15 \, \textrm{GeV}$] with 
$g_{\rho NN} = 3.25$ from the Bonn--J\"ulich model \cite{Machleidt:1987hj}.
Dotted line: Empirical spectral function from dispersion analysis
in the two--pion channel \cite{Belushkin:2005ds}.}
\label{fig:spect_f1}
\end{figure}

The one--loop chiral result for the isovector spectral function
near threshold can be stated as \cite{Strikman:2010pu}
\be
\frac{\textrm{Im} F_1^{V} (t + i0)}{\pi} 
&=& \frac{g_A^2 (t - 2 m_\pi^2)^2}{4 (4\pi f_\pi)^2 m_N \sqrt{t}}
(X - \arctan X) 
\phantom{xx}
\nonumber
\\
&+& \frac{2(1 - g_A^2) [k(t)]^3}{3 (4\pi f_\pi)^2 \sqrt{t}} ,
\label{imag}
\\[1ex]
X &\equiv& 4 m_N k(t) / (t - 2 m_\pi^2),
\label{imag_X}
\ee
where $g_A = 1.26$ is the nucleon isovector axial coupling, 
$f_\pi = 93\, \text{MeV}$ the pion decay constant,
and 
\be
k(t) &\equiv& \sqrt{t/4 - m_\pi^2} 
\label{k_cm}
\ee
the $t$--channel center--of--mass momentum of the $\pi\pi$ system
(here $t > 4\, m_\pi^2$).
Equations~(\ref{imag}) and (\ref{imag_X}) represent a compact 
approximation to the exact chiral 1--loop result with relativistic
nucleons \cite{Gasser:1987rb,Kaiser:2003qp} in which we omitted certain 
terms of order $t/m_N^2$ that become numerically important only at
$t \sim 1\, \textrm{GeV}^2$ and give negligible contributions to 
the charge density at large $b$.
In Fig.~\ref{fig:spect_f1} we compare our approximate expression
Eq.~(\ref{imag}) with the empirical spectral density obtained from 
the dispersion analysis of Ref.~\cite{Belushkin:2005ds}, which represents
an update of the classic result of Ref.~\cite{hoehler:book}. 
One sees that the one--loop result gives a reasonable representation 
of the empirical spectral density near threshold, with the discrepancy 
reaching $\sim 50\%$ at $t \approx 10\, m_\pi^2 = 0.195\, \textrm{GeV}^2$. 
Two--loop chiral corrections were studied in Ref.~\cite{Kaiser:2003qp}
and found to increase the value in this region by $\sim 20\%$.
We emphasize that the chiral expression is physically meaningful only in 
the near--threshold region $t - 4 m_\pi^2 \sim \textrm{few} \, m_\pi^2$; 
its numerical value at larger $t$ is shown for illustrative 
purposes only.
%
%

\begin{figure}
\includegraphics[width=0.48\textwidth]{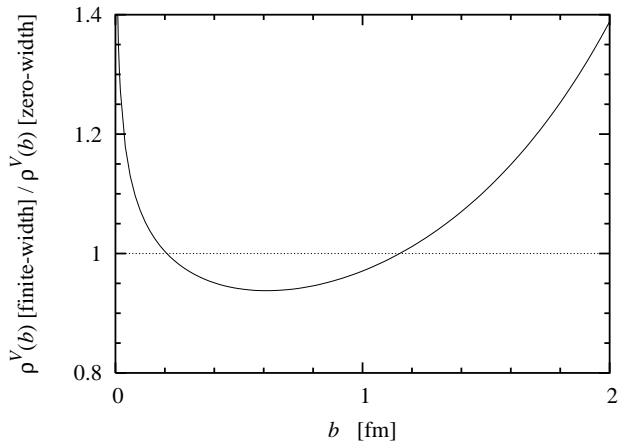}
\caption{Ratio of the isovector transverse charge densities $\rho^V(b)$
obtained from a finite--width $\rho$ meson [GS form
Eq.~(\ref{imag_gs}), $\Gamma_\rho = 0.15 \, \textrm{GeV}$] 
to the corresponding density for a zero--width pole. 
Both distributions are normalized to the same total charge.}
\label{fig:rhob_rat}
\end{figure}

In Ref.~\cite{Strikman:2010pu} we approximated the $\rho$ meson
contribution to the isovector charge density using a zero--width pole.
A simple theoretical model of the $\rho$ meson contribution
incorporating the finite $\rho$ width and its energy dependence
is the Gounaris--Sakurai (GS) form factor, obtained from an effective 
range expansion of the $\pi\pi$ scattering phase 
shift \cite{Gounaris:1968mw}. The spectral function 
of the resulting form factor $F_{\textrm{GS}}(t)$, normalized to
$F_{\textrm{GS}}(0) = 1$, is of the form
\be
\frac{\textrm{Im} \, F_{\textrm{GS}}(t + i0)}{\pi}
&=& \frac{C B(t)}{\pi [A^2(t) + B^2(t)]} ,
\label{imag_gs}
\ee
with
\be
A(t) &\equiv& m_\rho^2 - t + (\Gamma_\rho m_\rho^2/k_\rho^3) 
\{ k_\rho^2 h'_\rho (m_\rho^2 - t)
\nonumber \\
&+& [k(t)]^2 \, [h(t) - h_\rho] \} ,
\label{gs_A}
\\
B(t) &\equiv& (m_\rho^2 \Gamma_\rho /\sqrt{t} ) [k(t)/k_\rho]^3 ,
\label{gs_B}
\\
C &\equiv& m_\rho^2 + (\Gamma_\rho m_\rho^2/k_\rho^3)
[k_\rho^2 h'_\rho m_\rho^2 
\nonumber \\
&+& m_\pi^2 h_\rho - m_\pi^2 / \pi ] ,
\ee
where $\Gamma_\rho$ is the width parameter,
$k(t)$ the $t$--channel $\pi\pi$ center--of--mass momentum Eq.~(\ref{k_cm}),
$k_\rho \equiv k(m_\rho^2)$, and $h(t)$ denotes the auxiliary 
function
\be
h(t) &\equiv& \frac{2k(t)}{\pi \sqrt{t}} \, 
\ln \frac{\sqrt{t} + 2k(t)}{2 m_\pi} ,
\label{gs_h}
\ee
with $h_\rho \equiv h(m_\rho^2)$ and $h'_\rho \equiv dh/dt (m_\rho^2)$.
These expressions apply at $t > 4 m_\pi^2$. In fact, the full complex 
form factor on the upper edge of the cut at $t > 4 m_\pi^2$ is given by
\be
F_{\textrm{GS}}(t + i0) &=& \frac{C}{A(t) - i B(t)} ,
\label{gs}
\ee
and its values at $t < 4 m_\pi^2$ can be obtained by proper
analytic continuation of the expressions in Eqs.~(\ref{gs_A}),
(\ref{gs_B}), and (\ref{gs_h}). One finds that the form 
factor is regular at $t = 0$, as should be [the apparent singularity
from the $\sqrt{t}$ factors in Eqs.~(\ref{gs_B}) and (\ref{gs_h})
cancels between the two terms in the denominator of Eq.~(\ref{gs})]
and is normalized to unity there. In Fig.~\ref{fig:rhob_rat} we
compare the transverse charge density obtained from the finite--width
spectral function Eq.~(\ref{imag_gs}) et seq.\ 
($m_\rho = 0.77\, \textrm{GeV}, \Gamma_\rho = 0.15 \, \textrm{GeV}$)
with that obtained in the zero--width approximation; both densities
here are normalized to the same integral (total charge). One sees that
the zero--width form provides a very good approximation to the
charge density over a wide range of $b$, with an accuracy 
better than $10\%$ in the range 0.1-1.4 fm. At larger 
values of $b$ the finite--width density becomes systematically
larger than the zero--width approximation, reflecting the fact 
that large distances are dominated by the spectral strength 
at the lowest available masses.
%
%
%

%
%
\begin{figure}
\includegraphics[width=0.48\textwidth]{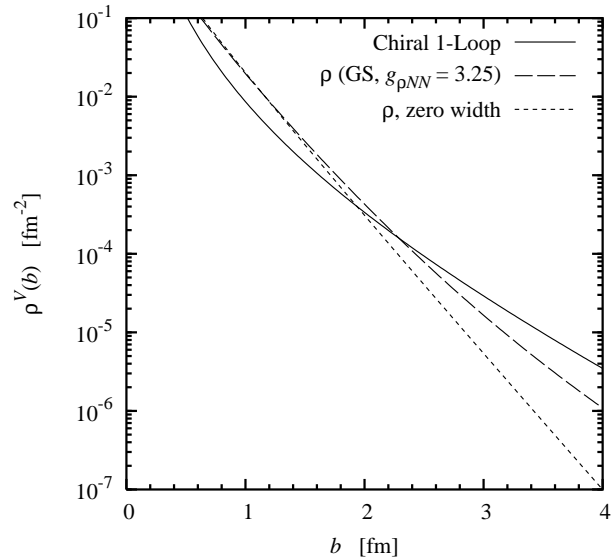}
\caption{Comparison of the transverse charge densities $\rho^V (s)$
resulting from the chiral and $\rho$ meson approximations
to the isovector spectral function (cf.\ Fig.~\ref{fig:spect_f1}). 
Solid line: Charge density from the one--loop chiral result
Eq.~(\ref{imag}). Dashed line: Charge density from a finite--width 
$\rho$ meson [GS form Eq.~(\ref{isovec_gs}), 
$\Gamma_\rho = 0.15 \, \textrm{GeV}$] with  $g_{\rho NN} = 3.25$ from the 
Bonn--J\"ulich model \cite{Machleidt:1987hj}.
Dotted line: Charge density from a zero--width $\rho$ meson pole
with the same coupling.}
\label{fig:rhob_chiral}
\end{figure}
The $\rho$ meson contribution to the isovector spectral function
in the GS approximation is then given by
\be
\frac{\textrm{Im} \, F^V_1 (t + i0)_{\rho}}{\pi} 
&=& \frac{g_{\rho NN}}{f_\rho}
\frac{\textrm{Im} \, F_{\textrm{GS}}(t + i0)}{\pi} ,
\label{isovec_gs}
\ee
where $g_{\rho NN}$ is the $\rho$--meson--nucleon vector 
coupling, and $f_\rho^{-1}$ parameterizes the $\rho$ meson coupling 
to the electromagnetic current and is related to the $e^+e^-$
partial decay width as $\Gamma (\rho \rightarrow e^+e^-) = 
(\alpha_{\rm em} m_\rho/3) (e/f_V)^2$. With the value of 
$g_{\rho NN} = 3.25$ from the Bonn--J\"ulich model of the
$NN$ interaction \cite{Machleidt:1987hj} and 
$f_\rho = 5.01$ from the experimental
value of the $e^+e^-$ partial decay width \cite{PDG} we
obtain
\beq
g_{\rho NN}/f_\rho \;\; = \;\; 0.65 .
\label{rho_coupling}
\eeq
This value is 30\% larger than the simple ``vector dominance'' value
of 0.5, which would follow from normalizing the $\rho$ contribution 
to the form factor given by Eq.~(\ref{isovec_gs}) to $F_1^V(0)_\rho = 1/2$,
and reflects the fact that in reality part of the charge carried
by the $\rho$ is compensated by the negative contribution from
higher--mass states. The spectral function resulting from a
$\rho$ meson with this coupling and a width $\Gamma_\rho = 0.15\, 
\textrm{GeV}$ is shown in Fig.~\ref{fig:spect_f1}. One sees that
this simple model agrees well with the empirical spectral function 
in the $\rho$ meson mass region.

Altogether, we see that the chiral component near threshold
and the finite--width $\rho$ meson with the above parameters
reproduce approximately the empirical isovector spectral function 
in the different regions. We emphasize that our aim here is 
not to construct a model of the complete spectral function, but 
merely to show that the theoretical approximations used in our earlier 
analysis \cite{Strikman:2010pu} agree reasonably well with the 
empirical result \textit{in their respective regions of validity.} 
In particular, we do not advocate to add the ``chiral''
and ``$\rho$'' components in Fig.~\ref{fig:spect_f1} (not 
even with an adjusted $\rho$ meson coupling), as this would imply 
that one has to evaluate the chiral expression in a region where 
it is not theoretically justified and compensate the result
by subtracting strength elsewhere; cf.\ the discussion in 
Ref.~\cite{Kaiser:2003qp}.

In Fig.~\ref{fig:rhob_chiral} we compare the chiral component of
the isovector charge density obtained from Eq.~(\ref{imag})
with that generated by the GS finite--width $\rho$ meson with 
the coupling Eq.~(\ref{rho_coupling}). It is seen that the chiral 
component dominates only at distances $b > 2 \, \textrm{fm}$.
An increase of the chiral component by $\sim 20\%$ due to two--loop 
corrections \cite{Kaiser:2003qp} would not substantially affect 
this comparison on a logarithmic scale. Also shown in 
Fig.~\ref{fig:rhob_chiral} is the density resulting from a zero--width 
$\rho$ meson pole of the same coupling, as used in the estimate of 
Ref.~\cite{Strikman:2010pu}. One notes that the finite width of
the $\rho$ meson pushes the region of dominance of the chiral
component out to even slightly larger distances.
Overall, the refined analysis here fully supports the conclusions
of Ref.~\cite{Strikman:2010pu}, that the chiral component of the
transverse charge density overwhelms the non--chiral density only
at distances $b > 2\, \textrm{fm}$.
\end{document}